\documentclass[aps,prl,twocolumn,superscriptaddress,floatfix,twoside]{revtex4}
\usepackage{amsmath}
\usepackage{amssymb}
\usepackage{graphicx}
\usepackage{dcolumn}
\usepackage{bm}

\mathchardef\ogon="012C
\newcommand{\as}{a\kern-0.22em\lower.40ex\hbox{$_{\ogon}$}}

\begin{document}

\title{Anomalous fermion bunching in density-density correlation}
\author{Hongwei Xiong\footnote{%
xionghongwei@wipm.ac.cn}}
\affiliation{State Key Laboratory of Magnetic Resonance and Atomic and Molecular Physics,
Wuhan Institute of Physics and Mathematics, Chinese Academy of Sciences,
Wuhan 430071, P. R. China}
\affiliation{Graduate School of the Chinese Academy of Sciences, P. R. China}
\author{Shujuan Liu}
\affiliation{State Key Laboratory of Magnetic Resonance and Atomic and Molecular Physics,
Wuhan Institute of Physics and Mathematics, Chinese Academy of Sciences,
Wuhan 430071, P. R. China}
\affiliation{Center for Cold Atom Physics, Chinese Academy of Sciences, Wuhan 430071, P.
R. China }
\affiliation{Graduate School of the Chinese Academy of Sciences, P. R. China}
\date{\today }

\begin{abstract}
We consider theoretically density-density correlation of identical Fermi
system by including the finite resolution of a detector and delta-function
term omitted in the ordinary method. We find an anomalous fermion bunching
effect, which is a quantum effect having no classical analogue. This
anomalous fermion bunching is studied for ultracold Fermi gases released
from a three-dimensional optical lattices. It is found that this anomalous
fermion bunching is supported by a recent experiment (T. Rom \textit{et al}
Nature 444, 733 (2006)).

\end{abstract}

\maketitle


Demonstrated firstly by Hanbury Brown and Twiss \cite{HBT}, quantum noise
correlation is a fundamental problem in identical Bose and Fermi system, and
has important application in astronomy, quantum optics, condensed matter
physics and subatomic physics \cite{Mandel,Scully,Baym,subatom} \textit{etc}%
. In the last ten years, quantum noise correlation was studied
experimentally for cold Bose and Fermi gases \cite%
{Shimizu,HBT-Cold,lat-bose,A-laser,Jin,Elonged,lat-Phi,lat-Fermi,Jeltes}
with the remarkable development of cold atom physics. Together with these
experimental advances, intensive theoretical studies contribute largely to
our understanding of correlation effect for cold atomic system. The
theoretical studies about high-order correlation have extended from a single
harmonically trapped Bose gas \cite{Glauber} to different matter state such
as pair correlations of fermionic superfluid \cite{Lukin,Strin} and
different trapping potential such as high-order correlation of cold atoms in
an optical lattice \cite{lattice-theo}. The finite-temperature effect \cite%
{Fin-tem} and low-dimensional effect \cite{dynamics-1D}\ about high-order
correlation were also studied theoretically.

The theoretical studies for fermionic superfluid \cite{Lukin,Strin} and
one-dimensional ultracold Fermi gas \cite{Mathey} have shown that quantum
noise correlation provides important information about the fundamental
quantum features of ultracold Fermi gases. Most recently, there is a clear
observation of fermion antibunching in a degenerate Fermi gas released from
a three-dimensional optical lattice \cite{lat-Fermi}. The fermion
antibunching physically arises from the anticommutation relation of field
operators, and thus has no classical analogue. It is a manifestation of
Pauli exclusion principle in high-order correlation. Besides the observation
of the antibunching effect in density-density correlation for fermionic
atoms released from an optical lattice \cite{lat-Fermi}, fermion
antibunching was also observed for electrons \cite%
{electron,electron1,electron2} and neutrons \cite{neutron}.

In the present Letter, we find that when the resolution of a detector for
the measurement of density distribution is considered, there would be an
anomalous fermion bunching effect under appropriate conditions. This
anomalous fermion bunching effect originates physically from the delta
function in the anticommutation relation of fermion field operators, omitted
in the ordinary method in calculating the density-density correlation. The
consideration of finite resolution of the detector will avoid the notorious
divergence in the delta-function term, and lead to anomalous bunching
effect. The observation of obvious anomalous fermion bunching effect
requests special conditions and parameters. Fortunately, we notice that
there is already an anomalous fermion bunching in the experimental data of
Ref. \cite{lat-Fermi}.

Because we will give a comparison between our theory and experimental data
to support the anomalous fermion bunching, we first introduce briefly the
elegant experiment in Ref. \cite{lat-Fermi}. In this experiment, an
ultracold Fermi gas of $^{40}\mathrm{K}$ was firstly prepared in the
combined potential of an optical trap and a 3D (three-dimensional) optical
lattice. After free expansion of $10$ \textrm{ms} by suddenly switching off
the combined potential, 2D (two-dimensional) density images were recorded
with a CCD (charge-coupled device) camera by illuminating a resonant laser
along the vertical ($z$) direction. See Fig. 1(a). The 2D density-density
correlation was obtained by dealing with appropriately a set of 2D density
images.

The starting point of our theory is the 2D density-density correlation
function given by%
\begin{equation}
C_{2}(\mathbf{d},t)=\frac{\int \langle \left\langle \widehat{n}(\mathbf{r-d}%
/2,t)\widehat{n}(\mathbf{r+d}/2,t)\right\rangle \rangle _{d}d^{2}\mathbf{r}}{%
\int \langle \left\langle \widehat{n}(\mathbf{r-d}/2,t)\right\rangle \rangle
_{d}\langle \left\langle \widehat{n}(\mathbf{r+d}/2,t)\right\rangle \rangle
_{d}d^{2}\mathbf{r}}.  \label{density Correla}
\end{equation}%
Here $\widehat{n}$ is a 2D density operator. $\mathbf{r}\equiv \{x,y\}$ and $%
\mathbf{d}\equiv \{d_{x},d_{y}\}$. To deal with appropriately the layered
distribution of fermionic atoms in 3D optical lattice and absorption imaging
along $z$ direction, the 2D density operator takes the form $\widehat{n}(%
\mathbf{r},t)=\widehat{g}\widehat{\Psi }^{\dag }(\mathbf{r},t)\widehat{\Psi }%
(\mathbf{r},t)$. The operator $\widehat{g}$ has the property that $%
\left\langle f(\widehat{g})\widehat{a}_{ij}^{\dag }\widehat{a}%
_{ij}\right\rangle =f(g_{ij})$. $g_{ij}=\sum_{k_{z}}f_{ijk_{z}}$ with $%
f_{ijk_{z}}$ being the occupation number in the lattice site indexed by%
\textit{\ }$\{i,j,k_{z}\}$. The summation is about the lattice site along $z$
direction. $g_{ij}$ represents the overall particle number in a string of
lattice sites along $z$ direction. $\widehat{a}_{ij}$ is an annihilation
operator for the atom in an equivalent 2D lattice site indexed by $\{i,j\}$.
The 2D density-density correlation function is dependent on $\mathbf{d}$. $%
C_{2}(\mathbf{d},t)<1$ corresponds to the fermion antibunching effect, while
$C_{2}(\mathbf{d},t)>1$ corresponds an anomalous fermion bunching effect.

We give here a brief introduction about the reason for the above rules in
the transformation from 3D system to 2D density-density correlation measured
by a CCD. Because the density images were recorded in Ref. \cite{lat-Fermi}
with a CCD camera by illuminating a resonant laser along $z$ direction,
special consideration should be given about the 2D density-density
correlation for a 3D system. The 2D density distribution is%
\begin{eqnarray}
n(\mathbf{r},t) &=&\int dz\left\langle \widehat{\Psi }^{\dag }(\mathbf{r}%
,z,t)\widehat{\Psi }(\mathbf{r},z,t)\right\rangle  \notag \\
&=&\Sigma _{ij}g_{ij}\left\vert \phi _{ij}(\mathbf{r},t)\right\vert ^{2}.
\label{2D}
\end{eqnarray}%
Here $\phi _{ij}(\mathbf{r},t)$ is the wave function of the atom initially
in the lattice site $\left\{ i,j\right\} $. Because in 2D density-density
correlation, the coordinate $z$ has been integrated, it is convenient to
define the 2D density operator as%
\begin{equation}
\widehat{n}(\mathbf{r},t)=\widehat{g}\widehat{\Psi }^{\dag }(\mathbf{r},t)%
\widehat{\Psi }(\mathbf{r},t).
\end{equation}%
With the above rules about $\widehat{g}$, $\left\langle \widehat{n}(\mathbf{r%
},t)\right\rangle $ is the same as the result given by equation (\ref{2D}).
With this definition of 2D density operator and $\widehat{g}$, it is
straightforward to get the final result of the 2D density-density
correlation function.

From Eq. (\ref{density Correla}), it is easy to get%
\begin{eqnarray}
&&\left\langle \widehat{n}(\mathbf{r}-\mathbf{d}/2,t)\widehat{n}(\mathbf{r}+%
\mathbf{d}/2,t)\right\rangle   \notag \\
&=&\left\langle \widehat{g}^{2}\widehat{\Psi }^{\dag }(\mathbf{r}-\mathbf{d}%
/2,t)\widehat{\Psi }(\mathbf{r}+\mathbf{d}/2,t)\right\rangle \delta (\mathbf{%
d})  \notag \\
&&+\left\langle \widehat{n}(\mathbf{r}-\mathbf{d}/2,t)\right\rangle
\left\langle \widehat{n}(\mathbf{r}+\mathbf{d}/2,t)\right\rangle   \notag \\
&&-\left\vert \left\langle \widehat{g}\widehat{\Psi }^{\dag }(\mathbf{r}-%
\mathbf{d}/2,t)\widehat{\Psi }(\mathbf{r}+\mathbf{d}/2,t)\right\rangle
\right\vert ^{2}.  \label{d-d Cor}
\end{eqnarray}%
The delta-function term in the above formula is ordinarily omitted \cite%
{lat-Fermi,Wick}\ because the divergent property and $\delta (\mathbf{d})=0$
for $\mathbf{d}\neq \mathbf{0}$. This term originates from the delta
function in the anticommutation relation $\left\{ \widehat{\Psi }(\mathbf{r}+%
\mathbf{d}/2,t),\widehat{\Psi }^{\dag }(\mathbf{r}-\mathbf{d}/2,t)\right\}
=\delta (\mathbf{d})$, and thus accounts for a pure quantum effect.

Bunching (antibunching) corresponds to a peak (dip) in density-density
correlation. For $\mathbf{d}\rightarrow \mathbf{0}$, the delta-function term
in equation (\ref{d-d Cor})\ means a divergent bunching behavior. When the
finite width $\Delta _{d}$ of spatial resolution is considered, it is
understandable that there would be an effect of increasing the width and
decreasing the height (depth) in the peak (dip) of the bunching
(antibunching) behavior. Roughly speaking, $\delta (\mathbf{d})$ may be
replaced by a function with height $1/\Delta _{d}^{2}$ in the region $%
-\Delta _{d}/2<d_{x}<\Delta _{d}/2$ and $-\Delta _{d}/2<d_{y}<\Delta _{d}/2$%
, and with zero value outside this region. When both the delta-function term
and spatial resolution are considered, more accurate results are obtained by
calculating%
\begin{equation*}
\langle \left\langle \widehat{n}(\mathbf{r}-\mathbf{d}/2,t)\widehat{n}(%
\mathbf{r}+\mathbf{d}/2,t)\right\rangle \rangle _{d}=
\end{equation*}%
\begin{equation}
\int_{\Xi }d^{2}\mathbf{s}_{1}\int_{\Xi }d^{2}\mathbf{s}_{2}\left\langle
\widehat{n}(\mathbf{r}-\mathbf{d}/2+\mathbf{s}_{1},t)\widehat{n}(\mathbf{r}+%
\mathbf{d}/2+\mathbf{s}_{2},t)\right\rangle .
\end{equation}%
Here $\Xi $ denotes the region $-\Delta _{d}/2<s_{x}<\Delta _{d}/2$ and $%
-\Delta _{d}/2<s_{y}<\Delta _{d}/2$. In the above formula, $\langle \rangle
_{d}$ represents the average due to the finite resolution of the detector.
This average has been considered in the starting point (\ref{density Correla}%
).

From equation (\ref{density Correla}), we get the following approximate
result%
\begin{equation}
C_{2}(\mathbf{d},t)\approx 1-\frac{\left\vert \Sigma _{jp}g_{jp}e^{-i2\pi
(d_{x}j+d_{y}p)/l}\right\vert ^{2}}{N^{2}}+\frac{Sl^{2}l_{p}^{2}\Sigma
_{ij}g_{ij}^{2}}{2\pi \Delta _{d}^{4}\sigma ^{2}N^{2}}.  \label{appro}
\end{equation}%
Here $l=2\pi \hbar t/ml_{p}$ with $m$ being the atomic mass and $l_{p}$
being the spatial period of the optical lattice. $S$ is the area of the
overlapping region between $\Xi _{1}$ (determined by $-\Delta
_{d}/2<x<\Delta _{d}/2$ and $-\Delta _{d}/2<y<\Delta _{d}/2$) and $\Xi _{2}$
(determined by $-\Delta _{d}/2<x-d_{x}<\Delta _{d}/2$ and $-\Delta
_{d}/2<y-d_{y}<\Delta _{d}/2$). $N$ is the total particle number. In
equation (\ref{appro}), the second term represents the antibunching
behavior. This term is obtained by omitting the effect of spatial resolution
and under the condition $l/\sigma >L_{0}/l_{p}$ and $l\gg \Delta
_{d}l_{p}/\sigma $. Here $L_{0}$ and $\sigma $ represent respectively the
overall width of the Fermi gas and wavepacket width of an atom in a lattice
site before switching off the combined potential. The last term in equation (%
\ref{appro}) physically originates from the delta-function term in the
anticommutation relation of field operators. This term is obtained by
considering the spatial resolution and under the condition $l/\sigma
>L_{0}/l_{p}$.

\begin{figure}[tbp]
\includegraphics[width=0.6\linewidth,angle=270]{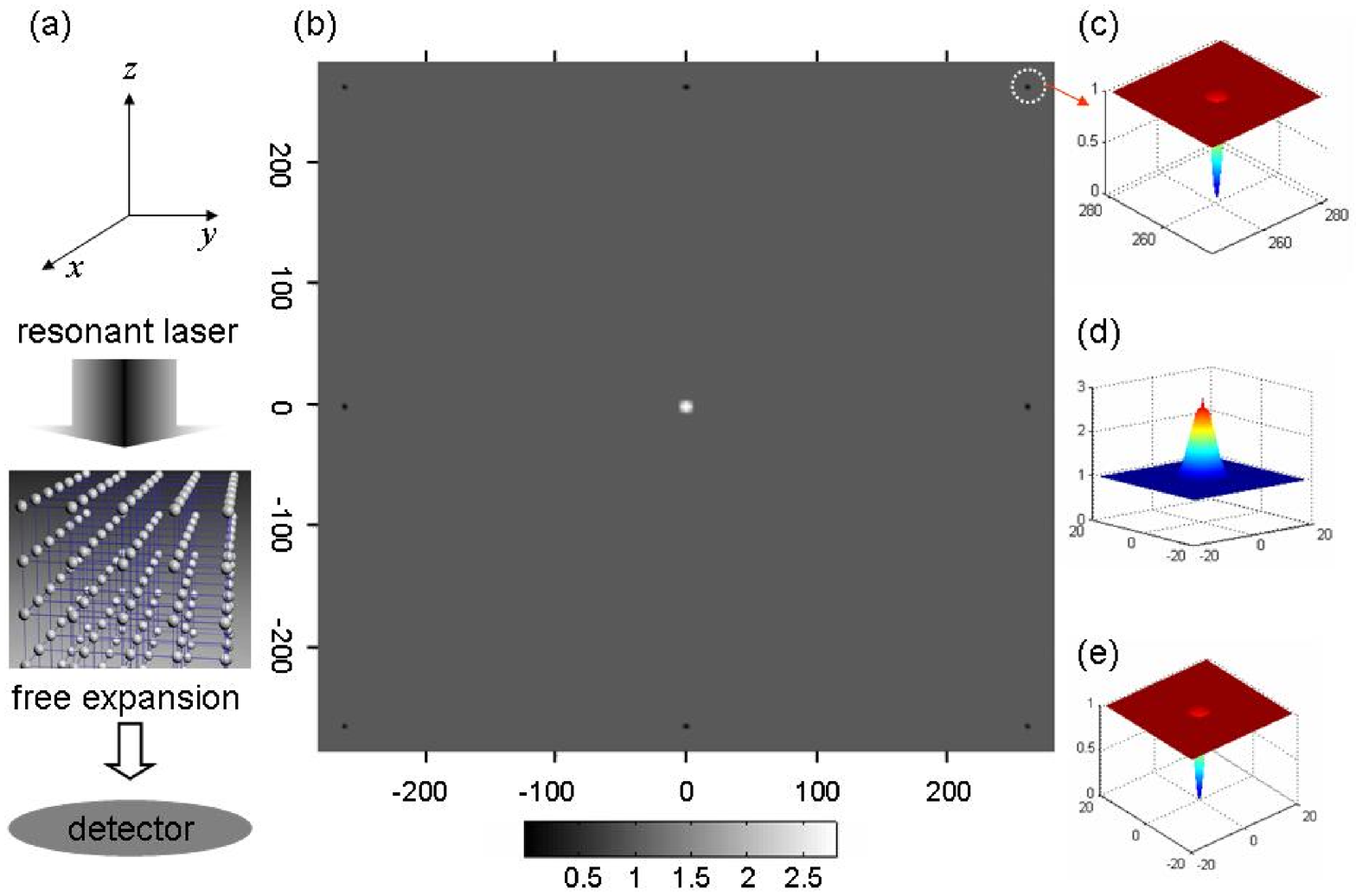}
\caption{Fermion antibunching and bunching in density-density correlation
function. (a) The measurement of density distribution for a Fermi gas
released from a three-dimensional optical lattice. (b) Two-dimensional
density-density correlation obtained from equation (\protect\ref{appro}) and
experimental parameters in Ref. \protect\cite{lat-Fermi}. The theoretical
results agree with the experimental data. In particular, an anomalous
fermion bunching is shown clearly by the bright dot at the center, which has
been supported strongly in Ref. \protect\cite{lat-Fermi}. (c) Illustration
of a dark dot (antibunching effect) in Fig.1b. (d) Illustration of the
bright dot (bunching effect). (e) The density-density correlation near the
center without the consideration of the delta-function term in the ordinary
theory. In these figures, the spatial coordinate is in unit of $\mathrm{%
\protect\mu m}$.}
\end{figure}

Using the experimental parameters in Ref. \cite{lat-Fermi} and equation (\ref%
{appro}), Fig.lb gives $C_{2}(\mathbf{d},t)$ for flight time of $10$ \textrm{%
ms}. Besides eight dark dots representing antibunching, at the center
location we notice a bright dot representing bunching. Fig.1c gives further
one of the dark dot, while Fig.1d gives further the bright dot. Eight dark
dots are due to the second term in equation (\ref{appro}). This term is $-1$
for $d_{x}/l=i$ and $d_{y}/l=j$ with $i$ and $j$ being integers. At the
locations of these dark dots, the last term in equation (\ref{appro}) is
zero. The bright dot at the center is due to the last term in equation (\ref%
{appro}), which is larger than $1$ close to $d_{x}=0$ and $d_{y}=0$. If the
last term in equation (\ref{appro}) is not included, there should be a dark
dot at the center (see Fig.1e). In Ref. \cite{lat-Fermi}, the theoretical
model without the last term is used to interpret their experimental data. It
is quite interesting to note that eight dark dots (rather than nine dark
dots without the consideration of the last term in equation (\ref{appro}))
were observed in Ref. \cite{lat-Fermi} (see Fig.2c and its figure caption in
this reference)! In Fig.2d of Ref. \cite{lat-Fermi}, the obvious and
\textquotedblleft mysterious\textquotedblright\ peak at the center of
density-density correlation shows further the anomalous fermion bunching
behavior. This comparison gives strong experimental evidence for anomalous
fermion bunching. It is clear that the condition for observing fermion
bunching is $l^{2}l_{p}^{2}\Sigma _{ij}g_{ij}^{2}>2\pi \Delta _{d}^{2}\sigma
^{2}N^{2}$.

The measurement of density-density correlation extracts the correlation
information in quantum noise. We discuss further the physical mechanism of
the bunching effect for Fermi system through the consideration of density
fluctuations. The density fluctuations are defined as%
\begin{equation}
\delta ^{2}n(\mathbf{x},t)=\lim_{\mathbf{y}\rightarrow \mathbf{x}%
}\left\langle \widehat{n}(\mathbf{x},t)\widehat{n}(\mathbf{y}%
,t)\right\rangle -\left\langle \widehat{n}(\mathbf{x},t)\right\rangle ^{2}.
\end{equation}%
Simple calculations give%
\begin{equation}
\delta ^{2}n(\mathbf{x},t)=\lim_{\mathbf{y}\rightarrow \mathbf{x}%
}\left\langle \widehat{\Psi }^{\dag }(\mathbf{x},t)\widehat{\Psi }(\mathbf{y}%
,t)\right\rangle \delta (\mathbf{x}-\mathbf{y})-\left\langle \widehat{n}(%
\mathbf{x},t)\right\rangle ^{2}.
\end{equation}%
We see that there is a divergent $\delta $-function term in $\delta ^{2}n(%
\mathbf{x},t)$. It is obvious that the omission of the $\delta $-function
term will lead to absurd result of negative density fluctuations. The
divergent $\delta $-function term is a pure quantum effect by noting that it
comes from the anticommutation relation between field operators. There is
another origin for this $\delta $-function term that the field operators $%
\widehat{\Psi }(\mathbf{x},t)$ and $\widehat{\Psi }^{\dag }(\mathbf{x},t)$
comprise infinite modes, rather than only the modes occupied by particles
before a measurement. In fact, this property is an essential property of
quantum field theory. The analyses about $\delta ^{2}n(\mathbf{x},t)$ show
clearly that the $\delta $-function term can not be omitted simply.
Similarly to our studies of the density-density correlation, the divergence
in the $\delta $-function term can be avoided because the resolution of a
detector always has a width. This is equivalent to the presentation that
creation or annihilation of a particle at an infinitesimal point is
impossible because this would mean an infinite energy exchange.

Note that the anomalous fermion bunching does not violate in any sense the
Pauli exclusion principle for identical fermions. The anomalous fermion
bunching originates from the delta-function term in the anticommutation
relation of field operators. Of course, the Pauli exclusion principle will
try to destroy the anomalous bunching effect. The second term in (\ref{appro}%
) reflects the Pauli exclusion principle, and leads to the antibunching
effect, and even can completely destroy the bunching effect at the center.

In summary, our studies give an anomalous fermion bunching, and we have
given a strong evidence by analyzing a recent experimental data in Ref. \cite%
{lat-Fermi}. The condition to observe the anomalous fermion bunching is in
fact quite rigorous. It is not surprising that this unique quantum effect is
found accidentally in Ref. \cite{lat-Fermi} with highly developed
experimental technique. The experimental technique in Ref. \cite{lat-Fermi}
gives us chance to test further our theory. (i) The last term in equation (%
\ref{appro}) increases with the increasing of flight time. Thus, we expect a
transition process from the antibunching to bunching at the center of $C_{2}(%
\mathbf{d},t)$, with the increasing of the flight time. (ii) The
distribution of fermionic atoms in the lattice sites $\{i,j,k_{z}\}$ is
determined by $i^{2}+j^{2}+\alpha _{z}^{2}k_{z}^{2}\leq R^{2}$ at zero
temperature. Here $\alpha _{z}$ is determined by the harmonic potential due
to the optical trap. Simple calculations give $\Sigma
_{ij}g_{ij}^{2}/N^{2}\simeq 0.93(\alpha _{z}N)^{-2/3}$. We see that
decreasing $\alpha _{z}$ has an effect of enhancing the anomalous fermion
bunching effect for identical particle number. This dependence on $\alpha
_{z}$ would give us further chance to test our theory. We believe further
experimental studies would deep largely our understanding of the
indistinguishability of identical particles, wave packet localization in
quantum measurement process and high-order correlation. High resolution of a
detector is required to reveal anomalous fermion bunching effect. The
astonishing resolution of STM (scanning tunneling microscope) and AFM
(atomic force microscope) \textit{etc} suggests that the experimental and
theoretical studies about electrons in an atom or molecule may lead to a new
regime about the studies of high-order correlation, quantum noise and
quantum measurement \textit{etc}.

The $\delta $-function term in the anticommutation relation is the direct
reason for the anomalous fermion bunching effect. For Bose system, the
inclusion of the $\delta $-function term in the commutation relation of
field operators may also play important role in density-density correlation.
It is easy to understand that the inclusion of the $\delta $-function term
will lead to an enhanced boson bunching effect at $\mathbf{d}=\mathbf{0}$.
In fact, in a recent experiment about the density-density correlation of
ultracold bosonic atoms released from an optical lattice, there is a clear
enhanced boson bunching effect at $\mathbf{d}=\mathbf{0}$ in $C_{2}(\mathbf{d%
},t)$ (see Fig. 2(c) in \cite{lat-bose}, where the brightness in the center
dot is much higher than other eight bright dots, and this can not be
explained without the consideration of the $\delta $-function term.). This
sort of enhanced boson bunching effect is also implied strongly in \cite%
{lat-Phi}. These analyses show that anomalous fermion bunching effect and
enhanced boson bunching effect have common physical origin---the
indistinguishability of identical particles and the $\delta $-function term
in the anticommutation (commutation) relation of\ field operators.

\begin{acknowledgments}
This work is supported by NSFC under Grant Nos. 10634060, 10474117 and NBRPC
under Grant Nos. 2006CB921406, 2005CB724508 and also funds from Chinese
Academy of Sciences.
\end{acknowledgments}

\end{document}